\documentclass[epj]{svjour}
\usepackage{latexsym}
\usepackage{graphicx}
\usepackage{amsmath}
\usepackage{amssymb}
\usepackage{xcolor}
\usepackage{ulem}

\begin{document}

\title{Non-Markovian dynamics in a spin star system: The failure of
  thermalization}

\author{ Zhihai Wang\inst{1} \and Yu Guo\inst{2} \and D.~L. Zhou\inst{1}}
\institute{Beijing National Laboratory for condensed matter
    physics, Institute of Physics, Chinese Academy of Sciences,
    Beijing, 100190, China \and School of Physics and
    Electronic Science, Changsha University of Science and Technology,
    Changsha 410114, China}
\date{ Received: date / Revised version: date}

\abstract
  {In most cases, a small system weakly interacting with a thermal bath
  will finally reach the thermal state with the temperature of the
  bath. We show that this intuitive picture is not always true by a
  spin star model where the non-Markov effect predominates in the
  whole dynamical process. The spin star system consists of a central
  spin homogeneously interacting with an ensemble of identical
  noninteracting spins. We find that the correlation time of the bath
  is infinite, which implies that the bath has a perfect memory, and
  that the dynamical evolution of the central spin must be
  non-Markovian. A direct consequence is that the final state of the
  central spin is not the thermal state equilibrium with the bath, but
  a steady state which depends on its initial state.
\PACS{
      {03.65.Yz}{quantum mechanics} \and
      {03.67.-a}{quantum information} \and
      {73.21.La}{electron states and collective excitation
      in quantum dot} }
}

\maketitle

\section{Introduction}

A quantum system in nature can never be completely isolated from its
surrounding environment, and in many cases it must be treated as an
open system~\cite{hp}.  With the recent progresses in
experiments, the investigations on quantum open systems have attracted
more and more attentions. Several relative theoretical approaches on
quantum open systems have been developed, such as the Markov
approximation~\cite{hp,cw,df}, the quantum jump
approach~\cite{hp04,wt}, the quantum state diffusion~\cite{dl,xy}, the
canonical transformation~\cite{xf,gj} as well as the path integral
method~\cite{legget}.

It is a common sense that, when a small system interacts with a large
thermal bath, it will finally reach the thermal state in equilibrium
with the bath. However, we will show that it is not always true by
studying the reduced dynamics of the central spin in a spin star
system where the central spin homogeneously interacts with an ensemble
of identical noninteracting spins. The spin star
configuration~\cite{farro1,hp04-1,yh,ah,mb,xxz,mbj,hp07,xz11,ef,yh09}
can be realized in many solid-state quantum systems, such as the
semiconductor quantum dot~\cite{lcy,wx06,wx07,wy08,wy09,wm} or the
nitrogen-vacancy center in diamond~\cite{nz12,nz11}.

In many studies, the environment interacting with a small system is
modeled as multimode bosons~\cite{legget} or spins~\cite{wa,hk}. For
such kind of environment, the correlation time of the environment is
often much shorter than the characteristic dynamic time of the
system. Especially, under the Markov approximation, the correlation
function is assumed to be a $\delta$ function of time
interval~\cite{hp,cw,df}, which implies that the environment's memory
effect can be neglected during the time evolution. However, in our
spin star configuration, the bath is composed of identical spins, and
the correlation function of the bath is found to be a periodic
oscillation function without damping. The thermal bath then has a
perfect memory and the reduced dynamics of the central spin must be
non-Markovian. Recently, the non-Markovian dynamics of the open system
is widely
investigated~\cite{ferro2,Palumbo,Kossakowski1,Kossakowski2,Kossakowski3,Kossakowski4},
and it shows that the backflow of the information occurs in the
non-Markovian process.

In literature, the exact reduced dynamics of the central spin in spin
star system has been studied. However, the free Hamiltonians of the
central spin and the thermal bath are neglected in
Refs.~\cite{hp04-1,yh} where the thermal effect of the bath can not be
discussed. Furthermore, the dephasing dynamics of the central spin is
investigated utilizing the time convolutionless (TCL) master equation
approach~\cite{dasarma1,dasarma2}, in which the TCL kernel can be
factorized.  Unfortunately, the analytical results can be obtained
only in the case of homogeneous Zeeman energies and isotropic
coupling. In the above papers, only the integrable systems, in which
the excitations are conserved, are involved.  In our work, we take
into account the free Hamiltonians and the anti-rotating wave terms in
the interaction, which lead the system non-integrable.  We find that
the dynamics is sensitive to the change of the temperature at low
temperatures, while it is nearly temperature independent at high
temperatures. Besides, the bath at high temperatures will induce a
significant decoherence.

In our system, the whole Hilbert space can be decomposed into a direct
sum of smaller subspaces due to the exchange symmetry of the bath
spins, so we are able to numerically study the dynamics of the system
exactly for up to $201$ bath spins. To describe the dissipation of the
central spin, we calculate the probability in its spin up state at
different bath temperatures. Our results show that, the probability
oscillates around a fixed value, and the oscillation amplitudes
decrease as the temperature rises. By altering the initial state of
the central spin, we find that the final state of the central spin is
a steady state dependent on its initial condition, instead of the
thermal equilibrium state as obtained under the Markov
approximation~\cite{gc}. The reason comes from the existence of the
invariant subspace during the evolution of our system.  The evolution
in one subspace is independent of that in another subspace, and the
central spin does not achieve equilibrium with the bath in any
subspace.  Therefore, the central spin can not be thermalized to
equilibrium with the spin bath.

The paper is organized as follows. In Sec.~\ref{model}, we set up our
model as a spin star configuration and investigate the reduced
dynamics of the central spin in detail. In Sec.~\ref{results}, we find
that the central spin can not reach the thermal state in equilibrium
with the bath, and we further analyze the underlying physical
mechanism with the aid of fast Fourier transformation. We also
briefly discuss the central spin decoherence at different temperatures. In
Sec.~\ref{markov}, we demonstrate that the traditional Markov
approximation is not applicable by investigating the correlation time
of the thermal bath as well as the mutual entropy between the central
spin and the thermal bath. In Sec.~\ref{conclusion}, we draw the
conclusions and make several remarks.

\section{The model and the reduced dynamics of the central spin}

\label{model}

\subsection{The Hamiltonian}

\begin{figure}[tbp]
\includegraphics[width=8cm]{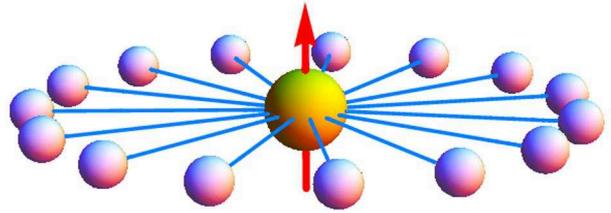}
\caption{(Color online) A schematic diagram of spin star
  configuration. The central spin with the energy level spacing
  $\protect\omega_0$ homogeneously couples to the surrounding bath
  spins with the energy level spacing $\protect%
  \omega$. The bath spins do not interact with each other.}
\label{bath}
\end{figure}

We consider a spin star configuration as shown in Fig.~\ref{bath}. The
system is composed of $N+1$ localized spin $1/2$ particles. The spin
labeled by the index $0$ locates at the center, and the surrounding
spins labeled by the index $i=1,2,\cdots, N$ are arranged in a circle.

The Hamiltonian of the global system is
\begin{equation}
  H=\frac{\omega _{0}}{2}\sigma _{z}^{(0)}+\frac{ \omega }{2}%
  \sum_{i=1}^{N}\sigma _{z}^{(i)}+g\sigma _{x}^{(0)}\sum_{i=1}^{N}\sigma
  _{x}^{(i)},  \label{hami}
\end{equation}
where $\sigma_{z}$ and $\sigma_{x}$ are the Pauli operators,
$\omega_{0}$ and $\omega$ are the frequencies of the central spin and
the bath spins respectively. The first two terms in the r.h.s. of
Eq.~(\ref{hami}) are the free Hamiltonians for the central spin and
the bath spins, while the last term describes that the central spin
interacts with the bath spins homogeneously and $g$ is the coupling
strength. Throughout this paper, we set $\hbar=1$.

Since the Hamiltonian (\ref{hami}) is invariant with respect to the exchange
of arbitrary two spins in the bath, it is reasonable to define the
collective operators:
\begin{equation}
  J_{z}=\sum_{i=1}^{N}\frac{\sigma _{z}^{(i)}}{2},J_{x}=\sum_{i=1}^{N}\frac{
    \sigma _{x}^{(i)}}{2}.  \label{col}
\end{equation}
Then the Hamiltonian (\ref{hami}) becomes
\begin{equation}
  H=\frac{\omega _{0}}{2} \sigma _{z}^{(0)}+\omega J_{z}+2g\sigma
  _{x}^{(0)}J_{x}.  \label{eq:ham}
\end{equation}
This implies that the central spin couples to a collective bath
angular momentum. Note that the angular momentum of the bath is a
changeable one and the total angular momentum $j$ can take values as
$j=N/2,N/2-1,\cdots,N/2-[N/2]$, where the notation $[a]$ denotes the
maximum integer not larger than $a$. Utilizing the exchange symmetry of the bath spins,
we are able to numerically investigate the reduced dynamics of the central
spin under the influence of the thermal bath exactly.

The state of the central spin is described by the reduced density matrix
\begin{equation}
  \rho ^{S}(t)=\mathrm{Tr}_{B}[\rho ^{SB}(t)],
\end{equation}
where $\rho^{SB} (t)$ is the density matrix of the global
system including the central spin and the thermal bath at time $t$
and $\mathrm{Tr}_{B}$ is the partial trace over the degree of freedom of
the thermal bath.  The density matrix $\rho ^{SB}(t)$ undergoes the
unitary evolution
\begin{equation}
  \rho^{SB} (t)=\exp (-iHt)\rho ^{SB}(0)\exp (iHt).
\end{equation}

In our consideration, the reduced density matrix of the central spin
$\rho^{S}$ is a $2\times 2$ matrix in the basis of the states $\left\{
  \left \vert \uparrow \right\rangle, \left\vert \downarrow
  \right\rangle \right\}$ with $\left\vert \uparrow \right\rangle$ and
$\left\vert \downarrow \right\rangle$ being the spin up and spin down
states respectively. The diagonal element $\left\langle \uparrow
\right. |\rho ^{S}|\uparrow \rangle $ corresponds to the probability
in its spin up state and the off-diagonal element $\left\langle \uparrow
\right. |\rho ^{S}|\downarrow \rangle $ or $\left\langle \downarrow
\right. |\rho ^{S}|\uparrow \rangle $ represents the coherence.

\subsection{The initial state of the thermal bath}

We assume that the initial state of the global system can be
factorized into an uncorrelated tensor product state:

\begin{equation}
\rho^{SB} (0)=\rho ^{S}(0)\otimes \rho ^{B}(0)  \label{eq:id}
\end{equation}
with $\rho ^{S}(0)$ and $\rho ^{B}(0)$ being the initial density matrices
for the central spin and the thermal bath respectively.

In our consideration, the initial state of the bath is a thermal equilibrium state
\begin{equation}
\rho ^{B}(0)=\frac{\exp (-\beta \omega J_{z})}{\mathrm{Tr}[\exp (-\beta
\omega J_{z})]},\label{initial_bath}
\end{equation}
where the $z$-component of the collective angular momentum $J_{z}$ is
defined in Eq.~(\ref{col}), and $\beta =1/k_{B}T$ is the inverse
temperature of the bath, with $k_{B}$ being the Boltzman constant,
which will be set to unit in the following.

The Hilbert space of the bath is an $N-$fold tensor product of two
dimensional spaces. A direct diagonalization of the Hamiltonian is
practically impossible when the number of bath spins $N$ is very
large.  Fortunately, the total angular momentum of the thermal bath is
a conserved quantity, i.e., $[J^{2},H]=0$. Due to the different possible orientations of the
spins in the thermal bath, the subspaces with total angular momentum $j$ has a degeneracy of
\begin{equation}
\alpha _{j}^{N}=\left(
\begin{array}{c}
N \\
\frac{N}{2}-j%
\end{array}%
\right) -\left(
\begin{array}{c}
N \\
\frac{N}{2}-j-1%
\end{array}%
\right).
\end{equation}
Therefore,  we introduce another index $\gamma_{j}=1,2,\cdots%
\alpha_j^{N}$
to label the
different degenerate subspaces sharing a same $j$.
 Thus the huge Hilbert space
$\mathbb{C}^{B}$ can be decomposed into a direct sum of smaller
invariant subspaces $\mathbb{C}^{B}_{j\gamma_j}$.  i.e.,
\begin{equation}
\mathbb{C}^{B}=\oplus_{j\gamma_j}\mathbb{C}^{B}_{j\gamma_j}.  \label{spa}
\end{equation}
Here, the $2j+1$ dimensional subspace $\mathbb{C}^{B}_{j\gamma_j}$ is
spanned by the states with the total angular momentum $j(j\leq N/2).$

The above analysis naturally motivates us to rewrite the initial
density matrix of the thermal bath $\rho ^{B}(0)$ in the form of a
direct summation.  To this end, we define a set of orthogonal bases
$|\gamma _{j},j,m\rangle$ in the subspace $\mathbb{C}_{j\gamma
  _{j}}^{B}.$ These states are the eigenstates of $J^{2}$ with
eigenvalues $j\left( j+1\right) $ and of $J_{z}$ with eigenvalues
$m$. Then, $\rho ^{B}(0)$ can be written as
\begin{eqnarray}
\rho ^{B}(0) &=&\frac{1}{Z}\exp (-\beta \omega J_{z})  \notag
\label{eq:rhob0} \\
&=&\frac{1}{Z}\sum_{j}\sum_{m=-j}^{j}\sum_{\gamma _{j}=1}^{\gamma
_{j}=\alpha _{j}^{N}}\exp (-\beta \omega m)|\gamma _{j},j,m\rangle \langle
\gamma _{j},j,m|,  \notag \\
&&
\end{eqnarray}%
where the partition function $Z$ is defined as
\begin{equation}
Z\equiv \mathrm{Tr}\big[ \sum_{j}\sum_{m=-j}^{j}\sum_{\gamma
_{j}=1}^{\gamma _{j}=\alpha _{j}^{N}}\exp (-\beta \omega m)|\gamma
_{j},j,m\rangle \langle \gamma _{j},j,m|\big] .  \label{par}
\end{equation}

\subsection{The reduced dynamics of the central spin}

In this subsection, we will study the reduced dynamics of the central spin
under the influence of the thermal bath.

As discussed in the previous subsection, the Hilbert space of the
thermal bath is composed of a direct sum of smaller invariant
subspaces.  Furthermore, the Hilbert space $\mathbb{C}^{SB}$ of the
global system (including the central spin and the thermal bath) is
given by $\mathbb{C}%
^{SB}\mathbb{=\oplus }_{j \gamma_j}\mathbb{C}_{j \gamma_j}^{SB},$
where
\begin{equation}
\mathbb{C}_{j\gamma_j}^{SB}\mathbb{=C}^{S}\otimes\mathbb{C}_{j \gamma_j}^{B}
\end{equation}
with $\mathbb{C}_{j\gamma_j}^{B}$ being defined in Eq.~(\ref{spa}) and
$%
\mathbb{C}^{S}$ being the $2$-dimensional Hilbert space for the
central spin. In the basis of
$|\mu;\gamma_{j},j,m\rangle\equiv|\mu\rangle\otimes|%
\gamma_{j},j,m\rangle$, where $|\mu\rangle=|\uparrow(\downarrow)\rangle$ is the
state of the central spin and $|\gamma_{j},j,m\rangle$ is
the state of the bath, the Hamiltonian is written as $H=\sum_{j} \sum
_{\gamma_{j}=1}^{\alpha_{j}^N} \mathcal{H}_{j\gamma_j}$ with
\begin{eqnarray}  \label{hj}
\mathcal{H}_{j\gamma_j}=\sum_{\mu,\mu^{\prime }=\downarrow ,\uparrow
}\sum_{m,m^{\prime }=-j}^{j} H_{\mu,m}^{\mu^{\prime },m^{\prime }}
|\mu^{\prime };\gamma_{j},j,m^{\prime }\rangle \langle \mu;\gamma_{j},j,m|
\notag \\
\end{eqnarray}
where the matrix element $H_{\mu,m}^{\mu^{\prime },m^{\prime }}$ is defined
as
\begin{equation}
H_{\mu,m}^{\mu^{\prime },m^{\prime }} \equiv \langle \mu^{\prime
};\gamma_{j},j,m^{\prime }| H|\mu;\gamma_{j},j,m \rangle.  \label{element}
\end{equation}

Correspondingly, we rewrite the initial density matrix of the global system
as
\begin{equation}
\rho^{SB} (0)=\sum _{j} \sum_{\gamma_j=1}^{\alpha_{j}^{N}}
Z_{j\gamma_j}\rho _{j\gamma_j}^{SB}(0)/Z
\end{equation}
with
\begin{equation}
\rho _{j\gamma_j}^{SB}(0)=\frac{1}{Z_{j\gamma_j}}\rho ^{S}(0)\otimes
\sum_{m=-j}^{j}\exp (-\beta \omega m)|\gamma_{j},j,m\rangle \langle
\gamma_{j}, j,m|
\end{equation}
where
\begin{equation}
Z_{j\gamma_j}=\mathrm{Tr}\big[\sum_{m=-j}^{j}\exp (-\beta \omega
m)|\gamma_{j},j,m\rangle \langle \gamma_{j}, j,m|\big]
\end{equation}
is introduced to guarantee the unity trace of
$\rho_{j\gamma_j}^{SB}(0)$. In the above equations,
$\mathcal{H}_{j\gamma_j}$ and $\rho _{j\gamma_j} ^{SB}(0)$ can be
understood as the projection of the Hamiltonian and the initial
density matrix on the subspace $\mathbb{C}_{j\gamma_j}^{SB}$
respectively.

In the subspace $\mathbb{C}_{j\gamma_j}^{SB},$ the ``density matrix'' $\rho
_{j\gamma_j}^{SB}$ experiences the unitary evolution
\begin{equation}
\rho _{j\gamma_j}^{SB}(t)=\exp (-i\mathcal{H}_{j\gamma_j}t)\rho
_{j\gamma_j}^{SB}(0)\exp (i\mathcal{H}_{j\gamma_j}t),
\end{equation}
and the reduced density matrix for the central spin is $\rho_{j%
\gamma_j}^{S}(t)=\mathrm{Tr}_{B}[\rho_{j\gamma_j}^{SB}(t)]$.

Hence the probability for the central spin in its spin up state is
\begin{equation}
P(t)=\frac{1}{Z}\sum_{j}\alpha _{j}^{N}Z_{j\gamma_j}\left\langle \uparrow
\right\vert \rho^{S} _{j\gamma_j}\left( t\right) \left\vert \uparrow
\right\rangle .  \label{eq:result}
\end{equation}
and the coherence function is
\begin{equation}
C(t)=\frac{1}{Z}\sum_{j}\alpha _{j}^{N}Z_{j\gamma_j}\left\langle \downarrow
\right\vert \rho^{S} _{j\gamma_j}\left( t\right) \left\vert \uparrow
\right\rangle .  \label{decoherence}
\end{equation}
\section{Results and discussions}

\label{results}
\subsection{the dissipation}
In general cases, a small system interacting with a large thermal bath
will be thermalized. In other words, the small system will reach the
thermal state in equilibrium with the bath. For example, if a single
spin-$1/2$ particle described by the free Hamiltonian
$H=\omega_{0}\sigma_{z}/2$
is taken as the small system, the final density matrix would be
\begin{equation}
  \rho(t=\infty)=\frac{\exp(-\beta \omega_{0} \sigma_{z}/2)}{\mathrm{Tr}%
    [\exp(- \beta \omega_{0} \sigma_{z}/2)]},  \label{estate}
\end{equation}
and the probability in its up state is
\begin{equation}
P(t=\infty)=\frac{1}{2}-\frac{\tanh(\beta \omega_{0}/2)}{2}.
\end{equation}
It shows that the probability depends only on the temperature of the
thermal bath but not on the initial state of the spin. Such results
can be obtained under the Markov approximation~\cite{gc,hefeng}. In
the following, however, we will show that this intuitive picture is
not true in our system.

\begin{figure}[tbp]
\includegraphics[width=8cm]{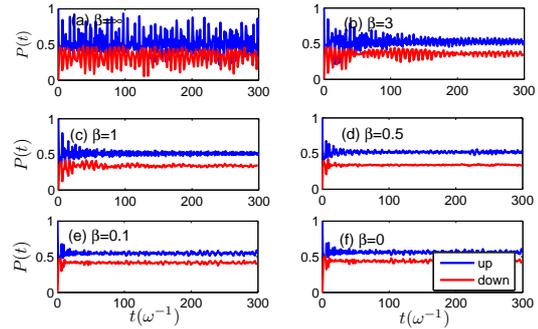}
\caption{(Color online) The probability for the central spin in its spin up
state as a function of time $t$. The blue upper(red lower) curve represents the case
when the central spin is prepared in its up (down) state initially. The
parameters are set as $\omega_{0}=1,g=0.1$. The inverse
temperature is (a)$\beta=\infty$, (b)$\beta=3$, (c)$\beta=1$,
(d)$\beta=0.5$, (e)$\beta=0.1$, (f)$\beta=0$. The number of spins in the
thermal bath is $N=201$. All parameters are in the unit of $\omega$.}
\label{dynam}
\end{figure}

Through a straightforward numerical calculation based on
Eq.~(\ref{eq:result}%
), we illustrate the probability of the central spin in its spin up
state as a function of the evolution time $t$ in Fig.~\ref{dynam}. In
the figure, the blue upper(red lower) curve represents the case when the central
spin is prepared in its up (down) state initially. Firstly, it shows
in the figure that the probability oscillates around a fixed value and
the amplitudes of the oscillation decrease with the increase of the
bath temperature. In other words, the probability tends to be a
constant when the temperature of the spin bath is high
enough. Secondly, the differences between the two
curves, especially the different steady values of the curves clearly
demonstrate that the final state of the central spin depends on the
temperature of the thermal bath as well as its own initial condition.

Furthermore, when the central spin is initially prepared in the spin
up or spin down state, we will have
$\langle\sigma_{x}^{(0)}\rangle=\langle%
\sigma_{y}^{(0)}\rangle=0$ during the time evolution. This can be
clarified in the viewpoint of $Z_2$ symmetry of the system. By the
analogy of Ref.~\cite{braak}, we define the parity operator
\begin{equation}
\mathcal{P}\equiv e^{%
  i\pi[J_z+\frac{\sigma^{(0)}_z}{2}+\frac{1+(-1)^{N}}{4}]}
\end{equation}
  where $N$ is the spin number
in the bath.  The parity is conserved with respect to the
Hamiltonian~(\ref{eq:ham}), namely, $[\mathcal{P},H]=0$. In our
consideration, the initial state has a fixed parity and it will stay
in the same parity during the time evolution due to the conservation
of the parity.  However, the central spin operators $\sigma_{x}^{(0)}$
and $\sigma_{y}^{(0)}$ change signs under the transformation of the
parity, i.e., $\mathcal{P}%
^{\dagger}\sigma_{x}^{(0)}\mathcal{P}=-\sigma_{x}^{(0)},\mathcal{P}%
^{\dagger}\sigma_{y}^{(0)}\mathcal{P}=-\sigma_{y}^{(0)}$. Therefore,
their average values will keep zero at any time $t$.

The above discussions show that, when the thermal bath is large
enough, the central spin will evolve into a state which satisfies
$\langle%
\sigma_{x}^{(0)} \rangle = \langle\sigma_{y}^{(0)}\rangle=0$ and $%
\langle\sigma_{z}^{(0)} \rangle \approx constant \neq \mathrm{Tr}
[\sigma_z^{(0)}\rho(t=\infty )]$.  Therefore, we can
safely conclude that the central spin will reach a steady state
different to the thermal equilibrium state. Furthermore, the steady
state depends on its initial state.

As demonstrated above, the central spin can not be 
thermalized as expected by interacting with an ensemble of identical
spins. This result can be understood as follows. The
initial thermal state of the bath has a distribution over the
subspaces and the energy level transitions in different subspaces
are independent of each other during the time evolution.  Even in an
arbitrary subspace, the central spin and the bath does not reach the
thermal equilibrium, and the time evolution in different subspaces are
independent of each other, so the global system can not achieve a thermal
equilibrium  state.

\begin{figure}[tbp]
\includegraphics[width=8cm]{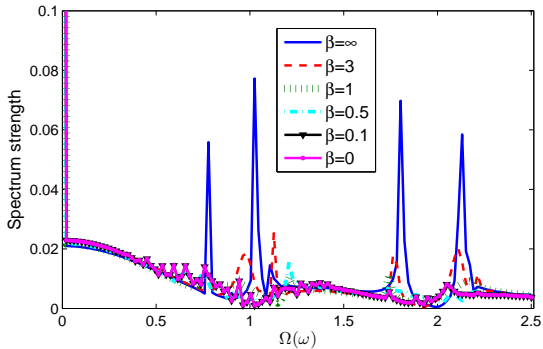}
\caption{(Color online) The frequency spectrum of the probability in Eq.~(%
\protect\ref{eq:result}) by FFT. The parameters are the same as that in Fig.~%
\protect\ref{dynam}. The central spin is in the spin up state initially.}
\label{spectrum}
\end{figure}

To investigate the energy level transitions in detail, we switch, by
the fast Fourier transformation (FFT) technology, from the time domain
to the frequency domain, and obtain the spectrum diagram which
manifests the frequency of the energy transition during time evolution
quantitatively.  We plot the frequency spectrum diagram for the
probability obtained from Eq.~(\ref{eq:result}) at different
temperatures in Fig.~\ref{spectrum} when the central spin is prepared
in its spin up state initially.

In zero temperature ($\beta=\infty$), all of the bath spins
are in their spin down states. In the condition of low excitations,
the Hamiltonian can be approximated as
\begin{equation}
H=\omega b^{\dagger}b+\frac{\omega_0}{2} \sigma_{z}+g\sqrt{N} \sigma_x ( b^{\dagger}+b)
\label{smode}
\end{equation}
where $b$ is defined as $b\equiv \sum_{i} \sigma_{i}^{-}/ \sqrt{N}$.
The central spin is then equivalent to coupling to
a single bath mode. Compared to the
standard Rabi model, which describes the interaction between a
two-level system and a single mode bosonic field~\cite{mo}, the
interaction has a $\sqrt{N}$ times enhancement. It is due to this
enhancement that the traditional rotating wave approximation loses its
effectiveness even in the case of $g\ll \omega,\omega_{0}$. In
Fig.~\ref{spectrum}, we obviously observe four distinct peaks. We note that
there will exist only one peak which corresponds to the Rabi frequency($2\sqrt{N}g$ in
our consideration). Besides, the inequality of the oscillation amplitudes in Fig.~\ref{dynam}(a) also
results from the effect of ``anti-rotating terms".

At low temperatures ($\beta=3,1$), the initial state only locates
in the subspaces $\mathbb{C}%
_{j\gamma_j}^{B}$ with large $j$, such as $j=N/2,N/2-1,N/2-2$.
It is shown that the peaks in the spectrum diagram and the probability
exhibit remarkable differences between the case of zero
temperature and low temperatures. Compared to the case of zero temperature,
the peaks and the oscillation amplitudes are lowered down dramatically
(shown in Figs.~\ref{dynam},~\ref{spectrum} ) at low temperatures.

As the temperature further rises ($\beta<1$), the components of
the initial state emerge in more and more subspaces
$\mathbb{C}^{B}_{j\gamma_j}$ with small $j$ and the relative weights
in the subspaces with large $j$ gradually decrease(shown in
Fig.~\ref{spectrum}). In these small $j$ subspaces, the energy level
differences are smaller than those in the large $j$
subspaces. Therefore, the high frequency peaks disappear while the
low frequency peaks arise (shown in Fig.~\ref{spectrum}). When the
temperature is high enough (for example when $\beta <0.1$), the frequency spectrum
and the dynamical evolution process are temperature independent. We see that the curves for
$\beta =0.1$ and $\beta =0$ coincide with each other in Fig.~\ref%
{spectrum} and Figs.~\ref{dynam}(e, f).

The frequency spectrum in Fig.~\ref{spectrum} shows that the
probability(the r.h.s. of Eq.(\ref{eq:result})) is a superposition of
oscillation functions with different frequencies and all of the
components evolve with time in unison during a short time. However,
the interference effect among different components becomes important once
the time is long enough, and then the probability oscillates around a fixed
value as time elapsed. The different behaviors of the probability
in two time scales are clearly shown at high temperatures in
Fig.~\ref{dynam} (d-f). That is, the probability monotonically decreases in a short time
and the revival occurs subsequently. We also find that the time when it shows revival decreases with
the increase of the number of spins in the bath(not shown in the figure).

\begin{figure}[tbp]
\includegraphics[width=8cm]{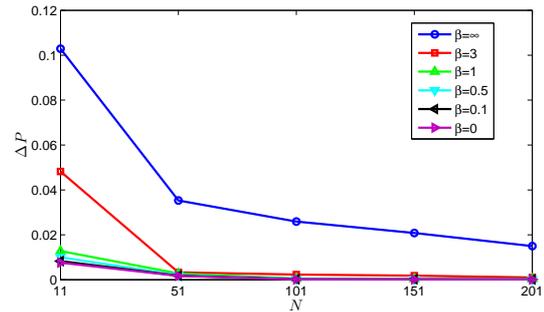}
\caption{(Color online) The fluctuation of the probability $\Delta P$ as a
function of the bath spins' number $N$ in different temperatures. The
parameters and initial state are same as Fig.~\ref{spectrum}. The curves for $%
\beta=0.1$ and $\beta=0$ completely coincide with each
other. }
\label{fluc}
\end{figure}

To describe the oscillation amplitudes of the probability
quantitatively, we define the fluctuation function
\begin{equation}
\Delta P\equiv\overline{\left( P(t)-\overline{P(t)}\right) ^{2}},
\end{equation}
where $\bar{a}$ denotes the mean value of $a$ over time. In
Fig.~\ref{fluc}, we plot the curve of $\Delta P$ as a function of the
bath spin number in different temperatures after the time when the
probability starts to oscillate around the fixed value (For example,
$t>50$ is taken in our consideration). It shows that $\Delta{P}$
decreases monotonously as the increase of the spin number and the
temperature of the bath. It implies that when the spin number in the
bath is large, the fluctuation function $\Delta P$ tends to zero at
high temperatures, and the probability will be a constant dependent of
the initial condition. Up to now, we reach the conclusion that the
steady state is only expected when the number of the bath spins $N$ is
infinity. For finite $N$, we will obtain a pseudo-stationary state at
high temperatures.

\subsection{The decoherence}
Decoherence refers to the process in which a quantum superposition
state is irreversibly transformed into a mixed state. In this section,
we will briefly discuss the decoherence of the central spin induced by
the coupling to the bath spins.

We prepare the central spin in the coherent superposition state
\begin{equation}
|\psi(0)\rangle=\frac{\left \vert\uparrow\right\rangle+\left \vert\downarrow\right\rangle}{\sqrt{2}}
\end{equation}
and the thermal bath in its thermal state which is described by the matrix density~(\ref{initial_bath}).
Then, the system will undergo the time evolution governed by the Hamiltonian~(\ref{hami}) or~(\ref{eq:ham}).
The coherence of the central spin at time $t$ is defined by
\begin{equation}
L(t)\equiv\frac{C(t)}{C(0)}
\end{equation}
where $C(t)$ is defined in Eq.~(\ref{decoherence}).
\begin{figure}[tbp]
\includegraphics[width=8cm]{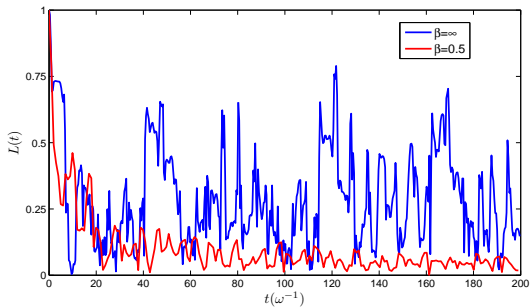}
\caption{(Color online) The central spin decoherence as a function of time
at different temperatures. The parameters
and the initial state are same as that in Fig.~\protect\ref{dynam}.}
\label{decoherence1}
\end{figure}
In Fig.~\ref{decoherence1}, we plot the central spin decoherence as a
function of time at different temperatures. It is shown that, at zero
temperature($\beta=\infty$), the coherence function $L(t)$ has a
obvious oscillation. However, at high temperature($\beta=0.5$), the
coherence function will have a minor oscillation nearby zero, which
means that the bath with high temperature induces the decoherence of
the central spin.

\section{The invalidity of the Markov approximation}

\label{markov}

In the study of the open system, the Markov approximation is widely
used to discuss the reduced dynamics of the system. The Markov
approximation works well when the small system weakly couples to a
large thermal bath which is composed of multimode bosons or spins and
the bath correlation function has a form as
$|\sum_{i}g_{i}^2e^{i\omega_i t}|$ where $\omega_i$ is the frequency
of the $i$th bath mode and $g_i$ its coupling strength to the
system. The sum of the oscillation functions with different
frequencies leads the bath correlation decaying more rapidly compared
to the system damping. As a result, the bath loses its memory and the
system does not affect the bath significantly, but the bath does
affect the system.  However, the Markov approximation loses its
effectiveness in our system even if the conditions of large thermal
bath and weak system-bath coupling are both satisfied. In this section
we will show the invalidity of the Markov approximation in two
aspects: the internal correlation time of the bath tends to be
infinite and the central spin does correlate with the bath all the
time during the time evolution.

Firstly, we discuss the correlation function of the thermal
bath. According to the standard master equation approach \cite{hp}, the
correlation function is calculated in the interaction picture with
$H_{0}=\omega_{0}\sigma_{z}^{(0)}/2+\omega\sum_{i}^{N}\sigma_{z}^{(i)}/2$,
and the result is expressed as
\begin{equation}
\langle \Gamma ^{\dagger }(t^{\prime })\Gamma (t)\rangle _{B}=g^{2}\frac{%
N\exp (-\beta \omega /2)}{2\cosh (\beta \omega /2)}e^{-i\omega (t-t^{\prime
})/2}  \label{eq:corr}
\end{equation}
with $\Gamma (t)=g\sum_{j}\sigma _{-}^{(j)}e^{-i\omega t}$ and
$\langle \cdot \rangle _{B}$ being the average over the thermal
equilibrium state of the bath which is described by
the density matrix
\begin{equation}
\rho^{B} =\frac{\exp \left( -\beta \omega \sum_{i}\sigma _{z}^{(i)}/2
\right) }{\mathrm{Tr}\left[ \exp \left( -\beta \omega /2 \sum_{i}\sigma
_{z}^{(i)}/2 \right) \right] }.
\end{equation}
Here, $\mathrm{Tr}$ means the trace over the degree of freedom of the
spin bath.

It is shown that the correlation function Eq.~(\ref{eq:corr}) is a
periodic oscillation function without any damping. Therefore, the
memory effect of the thermal bath can not be neglected and the
condition of the Markov approximation is naturally broken. We emphasis
that the non-decay of the correlation function results from the
homogeneous free terms of the bath spins, but not from the homogeneous
couplings.  From the viewpoint of the thermodynamics, the central spin
will still not be thermalized even in the inhomogeneous couplings.

Secondly, we will discuss the correlation between the central spin and
the thermal bath applying the concept of the mutual entropy. In
quantum information, the mutual entropy between two subsystems (the
central spin and the thermal bath in our system) is defined as
\begin{equation}
\mathcal{I}=S(\rho ^{S})+S(\rho ^{B})-S(\rho ^{SB})  \label{eq:me}
\end{equation}
where $S=-\sum_{i}\lambda_{i}\log_{2}\lambda_{i}$ ($\lambda_{i}$ being
the eigenvalues of $\rho $) is the von Neumann
entropy~\cite{aa,vv}. Roughly speaking, the mutual entropy
$\mathcal{I}$ quantifies how much information the central spin has
about the thermal bath and vice versa~\cite{mac}. In other words, the
mutual entropy characterizes in what extent the two subsystems
correlate with each other.

\begin{figure}[tbp]
\includegraphics[width=8cm]{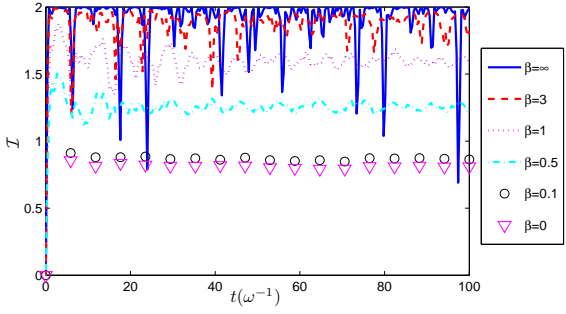}
\caption{(Color online) The mutual entropy between the central spin and the
thermal bath as a function of time at different temperatures. The parameters
and the initial state are same as that in Fig.~\protect\ref{dynam}.}
\label{entropy}
\end{figure}

Under the Markov approximation, the influence of
the open system to the bath is ignored, and the global system will
stay at the tensor product state
\begin{equation}
\rho^{SB} (t)=\rho ^{S}(t)\otimes \rho ^{B}_{th}
\end{equation}
with $\rho ^{B}_{th}$ being the initial thermal equilibrium state of
the bath. So, the mutual entropy between the open system and the
thermal bath disappears under the Markov approximation. However, this
is not the case for our system. In Fig.~\ref{entropy}, we plot the
mutual entropy between the central spin and the spin bath in our
system at different temperatures.  We observe that the mutual entropy
is much larger at low temperatures than high temperatures. Besides,
the entropy oscillates with the evolution time, and the oscillation
amplitudes become smaller as the increase of the temperature, which
exhibits the similar behavior as the probability discussed above. It
is shown in the figure that, the mutual entropy can reach its maximum
value $\mathcal{I}=2$ at low temperatures and a value little smaller
than $\mathcal{I}=1$ at high temperatures. In other words, the mutual
entropy never disappears in our system, which implies that the central
spin does correlate with the thermal bath, and is incompatible with
the results from the Markov approximation. This also verifies the
invalidity of the Markov approximation.

\section{Remarks and Conclusions }

\label{conclusion}
In this paper, we show that the central spin can not be perfectly
thermalized in the spin star system. We point out that, the
non-Markovian of the system depends on the peculiarity of the thermal
bath, since all the spins in the bath have an identical frequency, the
energy spectrum of the bath shows a high degeneration, and the
correlation function of the bath does not have a characteristic decay
time. This result, however, is independent of the initial state of the
central spin. On one hand, for the initial state satisfies
$\langle\sigma_z^{(0)}\rangle=\pm1$, the average values of
$\sigma_{x}^{(0)}$ and $\sigma_{y}^{(0)}$ keep zero during the time
evolution, which is directly proved from a consideration of parity
conservation, and the average value of $\sigma_{z}^{(0)}$ tends a
constant which is dependent of the initial state. On the other hand,
when the central spin is initially prepared in the state
$|\psi(0)\rangle=a\left\vert\uparrow\right\rangle
+b\left\vert\downarrow\right\rangle (|a|^2+|b|^2=1)$, we also observe
that the average values of $\sigma_{x}^{(0)}$ and $\sigma_{y}^{(0)}$
oscillate nearby zero, and the probability in its up state in this
case is the same as that for the initial state.
\begin{equation}
  \rho^{S}(0)=|a|^2\left\vert\uparrow \right\rangle  \left\langle \uparrow \right\vert+
  |b|^2\left\vert\downarrow \right\rangle  \left\langle \downarrow \right\vert.
\end{equation}
This observation also implies that the final state of the central spin
depends on its initial condition. In other words, the information of
the initial state is partially retained during the time evolution.

Technically, we point out that the Hilbert space is too huge to make a
direct numerical calculation in our system when the number of the bath
spins is very large. However, the huge Hilbert space can be decomposed into
a direct sum of smaller subspaces utilizing the exchange symmetry of
the bath spins, and we are still able to numerically study the reduced dynamics of
the system exactly when the environment is composed of hundreds of
spins. By using of the same approach, the entanglement and other
quantities in quantum information related to spin star system can be
solved without any approximation or specific numerical technique.

In summary, we study a non-integrable spin star system where the
Markov approximation is inapplicable for whatever weak coupling
between the system and the environment. In our system, a considerable
correlation between the system and the environment exists even in a
long time, and the central spin will finally evolve into a
pseudo-stationary state for a finite-size bath. We emphasize that the
final state for the central spin is dependent of its initial state,
i.e., the information about the initial state is partially
retained. We hope that our investigation on this simple system will
increase our understandings on the dynamics of quantum open system, a
central topic in the physical realizations of quantum information
processes.

\noindent \textbf{Acknowledgments.}
  We thank the discussions with C. P. Sun, D. Z. Xu and J. M. Zhang.
  This work was supported by National NSF of China (Grant
  Nos. 10975181, 11175247, and 11105020) and NKBRSF of China (Grant
  No. 2012CB922104). Guo Yu is also supported by the China
  Postdoctoral Science Foundation funded project.

\end{document}